\documentclass[acmsmall,screen]{acmart}

\renewcommand\footnotetextcopyrightpermission[1]{} 
\AtBeginDocument{%
  }

\usepackage[framemethod=TikZ, skipabove=10pt, roundcorner=10pt,innertopmargin=10pt,innerbottommargin=10pt]{mdframed}

\usepackage{hyperref}
\usepackage[inline]{enumitem}
\usepackage[textsize=footnotesize]{todonotes}
\setuptodonotes{inline}

\usepackage{xcolor}
\usepackage{pgfplots}
\usepackage{tikz}
\usepackage{multirow}
\usepackage{soul}
\usepackage{tabularx}

\definecolor{bblue}{HTML}{4F81BD}
\definecolor{rred}{HTML}{C0504D}
\definecolor{ggreen}{HTML}{9BBB59}
\definecolor{ppurple}{HTML}{9F4C7C}
\definecolor{yyellow}{HTML}{FFC300}

\usetikzlibrary{patterns}

\begin{document}

\title{Does Documentation Matter? An Empirical Study of Practitioners' Perspective on Open-Source Software Adoption}






\author{Aaron Imani}
\email{aaron.imani@uci.edu}
\orcid{0000-0001-7183-5468}
\affiliation{%
  \institution{University of California, Irvine}
  \city{Irvine}
  \state{California}
  \country{USA}
}

\author{Shiva Radmanesh}
\affiliation{%
  \institution{University of Calgary}
  \city{Calgary}
  \country{Canada}
  }
\email{shiva.radmanesh1@ucalgary.ca}

\author{Iftekhar Ahmed}
\affiliation{%
  \institution{University of California, Irvine}
  \city{Irvine}
  \state{California}
  \country{USA}
}
\email{iftekha@uci.edu}

\author{Mohammad Moshirpour}
\affiliation{%
  \institution{University of California, Irvine}
  \city{Irvine}
  \state{California}
  \country{USA}
}
\email{mmoshirp@uci.edu}

\renewcommand{\shortauthors}{Imani et al.}

\begin{abstract}
In recent years, open-source software (OSS) has become increasingly prevalent in developing software products. While OSS documentation is the primary source of information provided by the developers' community about a product, its role in the industry's adoption process has yet to be examined. We conducted semi-structured interviews and an online survey to provide insight into this area. 

Based on interviews and survey insights, we developed a topic model to collect relevant information from OSS documentation automatically. Additionally, according to our survey responses regarding challenges associated with OSS documentation, we propose a novel information augmentation approach, DocMentor, by combining OSS documentation corpus TF-IDF scores and ChatGPT. Through explaining technical terms and providing examples and references, our approach enhances the documentation context and improves practitioners' understanding. Our tool's effectiveness is assessed by surveying practitioners.
\end{abstract}

\ccsdesc[500]{Information systems~Open source software}
\ccsdesc[500]{Computing methodologies~Natural language processing}
\ccsdesc[500]{Software and its engineering~Documentation}

\keywords{open source, documentation, natural language processing, industry, interview}


\maketitle

\section{Introduction}


Open-source software (OSS) constitutes up to 95\% of codebases across various industrial sectors~\cite{Synopsis2023OpenReport}. Projections indicate that the OSS market value will surge from \$21.7 billion in 2021 to over \$50 billion in 2026, marking a remarkable 130\% increase~\cite{Rathee2022OpenTrends}. These statistics underscore a growing trend in the industry's adoption of OSS. However, OSS adoption carries significant risks when decisions are uninformed. For example, a design flaw in Log4j, a widely used OSS from Apache, led to a vulnerability allowing Remote Code Execution (RCE) attacks~\cite{Gupta2022IdentificationVulnerability}. This had repercussions on major software companies like Microsoft and Apple, emphasizing the critical role of reading and analyzing OSS documentation—the primary source of information for adoption and assessing OSS quality~\cite{Blumenthal2022ThinkingUp-to-Date}.


Practitioners base their adoption decisions on various information sources. In a recent survey conducted by Li et al. involving 23 participants from four different companies, five key sources of information in adoption decisions were identified~\cite{Li2022ExploringStudy}. These sources include version control systems, issue tracking systems, question and answer portals (Q\&A), forums and blogs, and security-related platforms. Notably, neither the survey participants nor the literature explicitly~\cite{Li2022ExploringStudy} mentioned OSS documentation as an information source during the Practitioners' Adoption Procedure (referred to as PAP). The absence of evidence regarding OSS documentation as a reference point and the absence of prior work addressing the identification of relevant OSS documentation types within PAP prompted the formulation of our initial research question.

\textbf{RQ1: Does OSS documentation play a role in practitioners' adoption procedure? What OSS documentation types are relevant to the adoption task?}

Similar to any decision-making scenario, choosing one OSS product (A) over another (B) relies on specific criteria. While software quality standards and OSS-specific quality models provide valuable evaluation criteria, it's important to note that prior research has revealed a limited industry inclination to utilize these readily available criteria. For instance, Yilmaz et al. conducted a systematic literature review covering 36 studies from 2003 to 2020 to identify quality evaluation models or frameworks for open-source software~\cite{Ylmaz2022QualityReview}. They found only one study that provided evidence, in the grey literature, of practitioners using quality evaluation models or frameworks to assess the quality of OSS within their development teams. However, achieving a comprehensive understanding of the role of OSS documentation in the PAP requires identifying the specific criteria practitioners seek in this resource throughout the adoption process. Thus, we outline our next research question.

\textbf{RQ2: What adoption criteria do practitioners look for in OSS documentation? Which criteria are most critical to them?}
Given the extensive categories within OSS documentation, we posit that not all information is crucial during the PAP. Practitioners should focus on identifying and extracting information that aligns with their adoption criteria from OSS documentation to make informed decisions. However, this task is not straightforward, as research suggests that developers' prior knowledge and cognitive biases can lead to inefficient and ineffective knowledge retrieval~\cite{deGraaf2014TheDocumentation}. Additionally, the information may not be easily discoverable, as documentation commonly suffers from usability issues related to information findability, where the intended information exists but users struggle to locate it~\cite{Aghajani2019SoftwareUnveiled}.



Consequently, the introduction of an automated mechanism aimed at extracting PAP-related information from OSS documentation can help mitigate the impact of practitioners' prior knowledge~\cite{deGraaf2014TheDocumentation} , which may render them susceptible to cognitive biases during PAP. This mechanism can accomplish this objective by discouraging practitioners from exclusively depending on the sections of OSS documentation they typically consult during PAP. Instead, providing them with a comprehensive catalog of pertinent information to review. Our next research question is prompted by this need.



\textbf{RQ3: How accurately can we identify the documentation topics related to the industry's selected criteria?}




Practitioners should fully comprehend relevant information from OSS documentation to make well-informed adoption decisions. Aghajani et al. mentioned the technicality level of the documentation as one of the readability issues of documentation \cite{Aghajani2019SoftwareUnveiled}. A search query of \textit{("jargon" or "too technical") and documentation} on GitHub returns hundreds of issues reported in OSS repositories (See \cite{kexkey2019,2019NeilHoskins,GH-nus-2020,2022shabir} for sample issues). OSS practitioners do not all possess the same domain knowledge in different OSS domains. Therefore, the provision of equal reading experiences for practitioners is critical to the success of PAP. This objective is the focus of our next research question.

\textbf{RQ4: How can we tailor OSS documentation for different levels of domain knowledge and reading goals?}

To answer RQs 1 and 2, we conducted semi-structured interviews with 10 practitioners with OSS adoption experience. To triangulate our interview results, we surveyed 42 practitioners. Following this, we created a corpus of OSS documentation to serve as the basis for addressing RQs 3 and 4. 
Since supervised deep learning relies heavily on label quality, we used the corpus to fine-tune a topic model called BERTopic \cite{Grootendorst2022BERTopic:Procedure} to address RQ3. 
Then, we developed our novel information augmentation approach, DocMentor, by combining OSS documentation corpus TF-IDF scores and ChatGPT \cite{openai}, a state-of-the-art large language model whose explanation capability has been successfully deployed in similar tasks \cite{Chen2023GPTutor:Explanation}, for answering RQ4. 
Lastly, we evaluated the usefulness and effectiveness of our solution by conducting follow-up survey with the individuals we interviewed initially.

The following are the key contributions of our study:

\begin{itemize}
    \item Providing the first OSS documentation corpus to enable further natural language processing research on OSS documentation.
    \item Identifying the role of OSS documentation in practitioners' adoption procedure.
    \item Collecting practitioners' adoption--relevant information from OSS documentation.
    \item Proposing a novel information augmentor that can not only enhance practitioners' understanding with any level of domain knowledge found in OSS documentation, but can also serve as a framework for constructing information augmentor applications across a variety of domains.
\end{itemize}

The remainder of this paper is arranged as follows: In Section \hyperlink{section.2}{2}, we review major research bodies related to our work. In Section \hyperlink{section.3}{3}, we outline our survey and interview methodology and discuss the obtained insights that help us answer RQs 1 and 2. Then, we present our steps to fine-tune a BERTopic topic model to answer RQ 3. In response to RQ 4, we introduce DocMentor in the final part of the section. The results of our topic model in Section \hyperlink{section.4}{4} are presented in terms of its accuracy in extracting PAP-related information.  Our evaluation survey to assess DocMentor's effectiveness and impact is discussed in Section \hyperlink{section.5}{5}. In Section \hyperlink{section.6}{6}, we analyze the threats to our methodologies' validity and the measures we followed to minimize them. In Section \hyperlink{section.7}{7}, our work is concluded, and future directions are outlined.

\section{Related Work}

Few studies have examined approaches to adopting OSS as components. Butler et al., \cite{Butler2022ConsiderationsBusinesses} interviewed 13 individuals from six Swedish software companies to examine contemporary approaches to adopting OSS as components, and the challenges encountered during this process. Based on participants' responses, the authors have categorized the challenges into four categories: technical aspects, license-related matters, OSS attributes, and risks. 
While Butler et al. investigated OSS quality factors, and adoption challenges, our study is the pioneering effort to specifically examine OSS adoption factors derived from OSS documentation, all within the context of \textit{industry practitioners' insights}

In a separate line of research, prior work has looked into automated ways of measuring the documentation quality. Tang et al.~\cite{Tang2023EvaluatingQuality} introduced an automated documentation quality evaluation tool, which involved the utilization of literature-derived metrics to assess documentation quality. 
Treude et al. have proposed a ten criterion framework to evaluate software documentation from different perspectives, such as structure, content, and style \cite{Treude2020BeyondQuality}. In particular, their framework identifies quality, appeal, readability, understandability, structure, cohesion, conciseness, effectiveness, consistency, and clarity as different aspects of software documentation quality. Carvalho et al. have proposed a tree-based toolkit called DMOSS to systematically assess the quality of non-source code text available in software packages \cite{Carvalho2014DMOSS:Assessment}. Utilizing the built-in plugins provided with DMOSS, they generate a report containing information about features such as the validity of the links available in the non-source code text, the validity of the license, spelling checks, the number of comments, source code coverage for documentation, and analysis of changes.



In order to investigate the importance of software documentation quality on software quality attributes and to identify the top important aspects of software documentation quality, Plösch et al. surveyed 88 participants working in the field of software engineering \cite{Plosch2014TheQuality}. According to their findings, the most influential quality attributes of software documentation are accuracy, clarity, readability, structuredness, and understandability. Moreover, the survey revealed that poor software documentation quality significantly affects analyzability. The aforementioned studies primarily center around software documentation quality. In contrast, our objective is to pinpoint the challenges faced by practitioners when utilizing OSS documentation throughout the PAP, with a specific focus on addressing the most commonly reported issues.

In recent years, there has been a growing body of research leveraging NLP techniques for various software engineering tasks. Ahmed et al. developed a framework for extracting and prioritizing software quality attributes in Agile-driven software development using Natural Language Processing (\textbf{NLP}) techniques \cite{Ahmed2023AnDevelopment}. 


To facilitate early architecture decision-making in agile software development, Gilson et al. used machine learning to extract software quality attributes defined in ISO/IEC 25010 quality models from user stories \cite{Gilson2019ExtractingMaking}. As a result of their analysis, they found that different precision and recall scores were obtained for classifying user stories based on each software quality attribute. Despite mentioning spaCy and pre-trained models, the authors did not go into detail about the machine learning model and manual features used in training the model.



Notably, our work is the first study to utilize an unsupervised deep-learning approach to extract PAP-relevant information from OSS documentation.

\section{Methodology}

\begin{figure}[tp]
  \centering
  \includegraphics[width=0.6\textwidth]{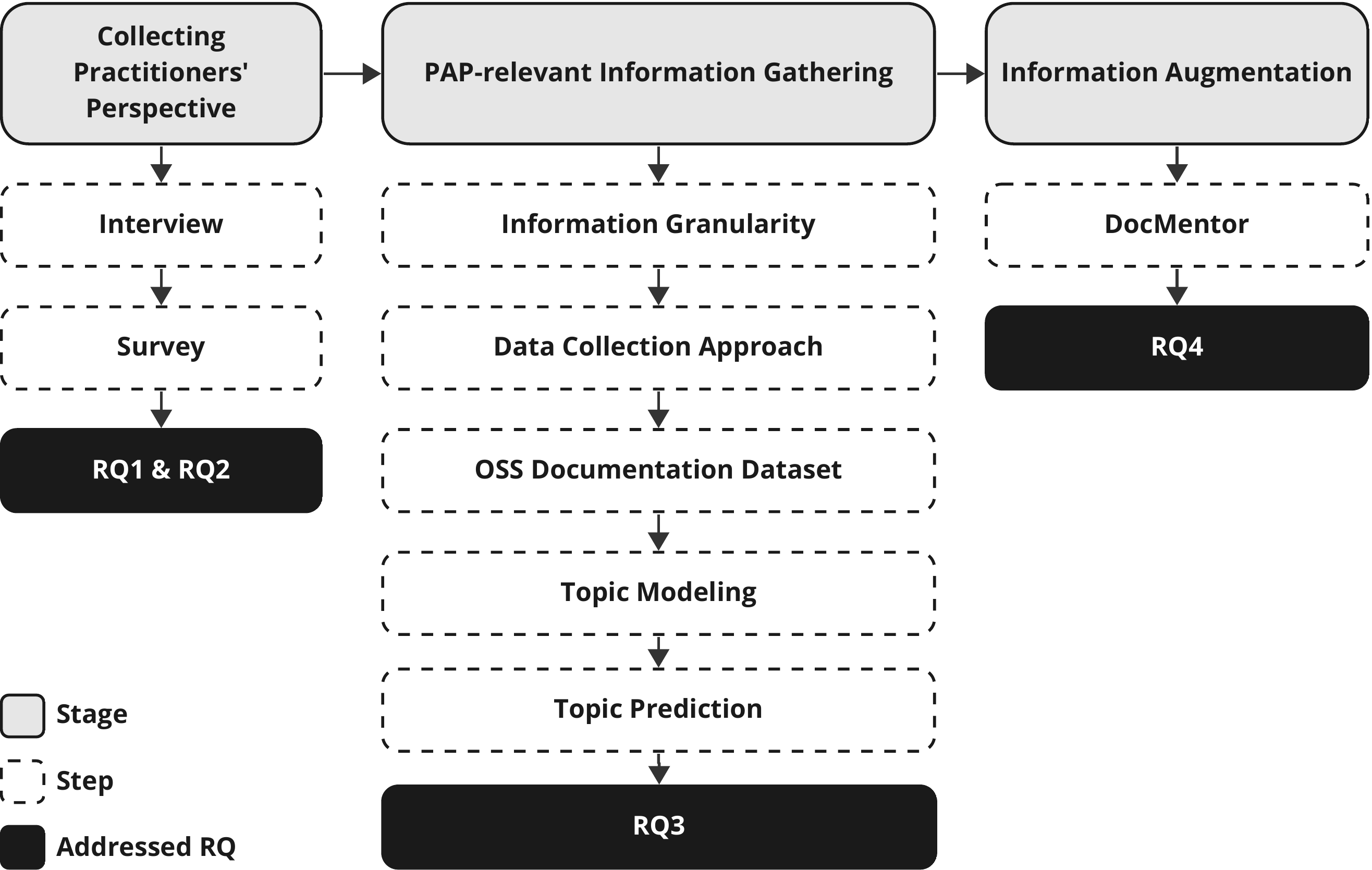}
  \caption{Overall methodology overview}
  \label{fig:meth-intro}
\end{figure}

In this section, we elaborate on our methodology to answer our RQs in three stages. In each stage, steps are taken to reach an answer for the corresponding stage's RQs. In the first stage, we gathered practitioners' perspectives through semi-structured interviews and an online survey to address RQ1 and RQ2. Following the RQ1 and RQ2 answers, we designed a solution to gather PAP-relevant information from OSS documentation to address RQ3 in the second stage. In the final stage, our proposed information augmentation approach, DocMentor, is presented. Our overall methodology is illustrated in Figure \ref{fig:meth-intro}, along with the steps involved in each stage.

\begin{table*}[bp]
\footnotesize
\setlength{\tabcolsep}{3pt}
  \caption{Interview participants' demographics}
  \label{tab:interview-participants}
  \begin{tabularx}{\textwidth}{XXXXl}
    \toprule
    ID&Experience in software industry (years)&Experience in OSS (years)&Company size&Role in company\\
    \midrule
    P1 & 21 & 21 & 85 & Senior software developer\\
    P2 & 17 & 17 & 200 & Technology principle\\
    P3 & 20 & 20 & 750 & Senior stack developer\\
    P4 & 6.5 & 6.5 & 750 & Senior software developer\\
    P5 & 32 & 20 & 50 & Chief engineer\\
    P6 & 34 & 34 & 1000 & Senior team lead\\
    P7 & 30 & 20 & 100 & Principal staff engineer\\
    P8 & 4 & 3 & 50,000 & Senior software engineer\\
    P9 & 22 & 22 & 20 & Chief technical officer\\
    P10 & 20 & 20 & 1,500 & Software engineering manager\\
  \bottomrule
\end{tabularx}
\end{table*}

\subsection{Interviews}
 We conducted semi-structured interviews with ten individuals from the software industry. These participants collectively possessed an average of 20 years of experience in integrating OSS into their company's products. The companies involved in the study represented a diverse range of sectors, including software development, software consulting services, software solutions for the oil and gas industry, cybersecurity, finance technology, legal services, AI-powered predictive tools for material and chemical companies, and engineering services. The interviews were conducted remotely, with an average duration of 30 minutes per session. We recorded the interviews with participants' consent, following the university-approved Institutional Review Board (IRB) protocol, and then transcripted them for analysis. Table \ref{tab:interview-participants} presents demographic information about our participants. 

During the interview, we asked specific questions related to the participants' utilization of OSS documentation during their decision-making process to adopt OSS projects into their products. These questions encompassed various aspects, including whether they rely on OSS documentation, the significance they attribute to documentation in their decision-making, the specific ways in which they utilize documentation during this process, the adoption criteria they prioritize while evaluating OSS documentation, their level of confidence in finding relevant information within OSS documentation that aligns with these criteria, challenges encountered when using OSS documentation, other textual resources they review when making decisions, how they utilize these resources, and the metrics they seek within them. See supplementary \cite{Imani2024DoesAdoption} for interview questions.


To enhance the depth of our interview data analysis, we categorized the responses provided by the interviewees. This was done for two main purposes. First, to facilitate the classification of various aspects of documentation into categories that have greater significance for the industry. Second, to employ these categories in the subsequent survey's multiple-choice questions. This approach allows for a more effective and consistent comparison between the outcomes of the interviews and the survey.

To accomplish this task, two authors independently reviewed the transcriptions and categorized participants' responses of the following questions:
\begin{enumerate*}[label=(\arabic*)]
    \item How do participants use OSS documentation while deciding to adopt an open-source project for integration into their products?
    \item What specific adoption criteria do participants seek while scrutinizing OSS documentation?
    \item What challenges do participants encounter when using OSS documentation?
    \item Apart from OSS documentation, what other textual resources do participants review when making decisions?
    \item How do they utilize these resources?
    \item What metrics do participants seek within these textual resources?
\end{enumerate*} Following individual categorization, the two authors engaged in negotiated agreement \cite{Forman2007QualitativeAnalysis, Garrison2006RevisitingReliability} on the categories for each question to reach a consensus. 
This resulted in the identification of nine distinct categories that participants consider when evaluating OSS documentation for adoption purposes. These categories are as follows:

\begin{enumerate}[label=\arabic*.]
    \item Project Maintenance
    \item Compatibility
    \item Functionality
    \item License Compatibility
    \item Community Adoption
    \item Existence of Examples of Use Cases
    \item Ease of Use
    \item Project's Versioning
    \item Performance
\end{enumerate}

\subsection{Online Survey}

To triangulate our findings and further validate the insights gathered from the interviews, we conducted a survey among practitioners within the software industry. Specifically, we targeted individuals who have first-hand experience in the process of adopting OSS for integration into their products. 

\subsubsection{Survey design}
Researchers conducted a pilot study and iteratively tuned the survey questions to ensure they were understandable and consistent with the survey objectives. The survey comprised a total of 24 questions, which were designed to gather insights from participants with diverse perspectives and experiences. The questions were a mix of multiple-choice, Likert scale, and open-ended questions, allowing us to capture both quantitative and qualitative data (See supplementary \cite{Imani2024DoesAdoption} for survey questions). The structure of the survey is as follows: Q3-Q6: Demographic questions. Q8-Q9: Questions assessing the participants' perception of the importance of OSS documentation in the OSS adoption decision-making process. Q10: A question about participants' approach to selecting an OSS for adoption. Q11-Q13, Q15: Questions exploring the factors influencing their decision to adopt an OSS. Q14: A question addressing the challenges encountered when using OSS documentation for making adoption decisions. Q15-Q20: Questions investigating the additional resources that participants refer to while making OSS adoption decisions and how they utilize these resources. Q21: A question focusing on the factors contributing to the ease of understanding one OSS documentation compared to another.

\subsubsection{Survey participants}
We utilized Qualtrics as our survey distribution platform. We targeted practitioners with prior experience in integrating OSS into their projects. We initially distributed the survey to our industry connections and followed up with periodic reminders. Additionally, we used the snowball sampling method \cite{Wohlin2014GuidelinesEngineering} by encouraging them to share the survey with other eligible employees within their respective companies. And eventually, we shared the survey on LinkedIn. The survey remained accessible for a duration of one month, during which we gathered a total of 29 responses.

\subsubsection{Participants' demographics}
Most of the respondents were affiliated with companies in the field of web development. Participants were also from open-source software, manufacturing, IoT, and finance companies. In terms of company size, 17.6\% had less than 50 employees, 25.71\% had 50-100 employees, 20\% had 100-500 employees, 8\% had 500-1000 employees, and 28.57\% had more than 1000 employees. 
On average, the participants had 11-15 years of experience in the software industry and 6-10 years of experience in integrating OSS into their projects.


\subsection{PAP-relevant Information Gathering}

In light of the findings from the survey and interviews, we systematically outline our methodology to address RQ3 in a step-by-step manner.

\subsubsection{Information Resource}
Given the diversity of open-source software (OSS) documentation, encompassing various platforms such as GitHub pages, Sphinx, MkDocs, and customized websites, our approach needed to narrow down to a specific documentation source. The selection criteria prioritized a unified structure both at the page level and throughout the documentation. 
Sphinx, which employs the reStructuredText (reST) markup language to create documentation in multiple formats, including HTML, provides a uniform format for OSS documentation \cite{sphinxdocs}. Its hierarchical structure and standardized format facilitate effective data extraction, contributing to the creation of our OSS documentation corpus and dataset.
Notably, Read the Docs hosts documentation for over 100,000 OSS projects \cite{readthedocs2023}, offering a substantial and valuable dataset for OSS documentation research. Furthermore, since it supports reStructuredText, it serves as a hosting platform for OSS projects employing Sphinx for documentation. As a result, to streamline the management of our OSS documentation collection, we focused on Sphinx-based documentation hosted by Read the Docs.

\subsubsection{Information Granularity}
Determining the suitable level of granularity for extracting PAP-relevant information from OSS documentation represents the initial step in this process. One potential approach involves inferring relevance from the web page titles of OSS documentation. This approach assumes that there is consistency in content topics within a webpage. However, OSS documentation webpages often contain multiple topics, especially in cases where poorly structured documentation combines incoherent information into lengthy web pages. Therefore, it becomes essential to pinpoint relevant information at a finer-grained level. Since our study is centered on Sphinx-based documentation
, we can establish the appropriate granularity by utilizing the information structure of reST. ReST organizes documents into sections, which subsequently translate into header tags in HTML output. Each section can encompass various types of content, including other sections. Consequently, the innermost sections (hereafter referred to as "sections") constitute the most cohesive units of information in Sphinx-based documentation. 
While one might contend that paragraphs offer an appropriate level of granularity, it is important to consider that paragraphs within a section often address specific subjects or topics as whole. Consequently, selecting only a subset of paragraphs from a section could result in incomplete coverage of the information presented within that section.


\subsubsection{Information Gathering Approach}
One approach for identifying PAP-relevant sections in OSS documentation is through the utilization of keywords and regular expressions. However, due to the wide range of OSS domains, including areas like deep learning and web development, domain-specific terminology can affect the completeness of our keyword coverage. Additionally, this approach overlooks content semantics, potentially resulting in the omission of sections that have indirect relevance to an adoption criterion. Therefore, it becomes imperative to develop an automated method that incorporates semantics and minimizes the need for human judgment, ultimately reducing subjectivity in the process.

To address these challenges, we turned to \textit{Topic modeling}, a well-established text analysis approach that has been in use for over two decades \cite{Zhao2021TopicSurvey}.
In recent years, the fusion of topic modeling with deep learning techniques has given rise to innovative topic models that surpass traditional Bayesian probabilistic models like Latent Dirichlet Allocation (LDA)~\cite{Blei2003LatentAllocation}. Some noteworthy examples of state-of-the-art topic models that rely on pre-trained embedding models include Top2vec~\cite{Angelov2020Top2vec:Topics}, CTM \cite{Bianchi2021Cross-lingualLearning}, and the latest addition, BERTopic~\cite{Grootendorst2022BERTopic:Procedure}.

Similar to Top2vec, BERTopic performs topic modeling by clustering document embeddings. However, unlike Top2vec, which represents topics based on cluster centroids, BERTopic uses a class-based version of TF-IDF to derive the topic representation from each topic. In comparative evaluations with LDA, Top2vec, and CTM, BERTopic has demonstrated state-of-the-art results on topic modeling benchmark datasets in terms of topic coherence and frequency~\cite{Grootendorst2022BERTopic:Procedure,Egger2022APosts}.
Consequently, we have chosen to employ BERTopic for extracting PAP-relevant information for two primary reasons: 1) It is an unsupervised technique that requires minimal human intervention, and 2) It captures the semantics of sections, which distinguishes it from keyword-based methods and traditional topic modeling.



\subsubsection{OSS Documentation Dataset}
Topic modeling with BERTopic necessitates a corpus of documents, which, in our case, needed to consist of OSS documentation. As there was no readily available corpus, we took the initiative to create one. Our rationale for this approach was to avoid potential bias towards terminology specific to a particular OSS domain. Consequently, we took measures to ensure diversity in the OSS domains represented within our corpus.

In a recent study by Sas et al., they introduced GitRanking, a taxonomy of software domains constructed from the ground up~\cite{Sas2022Gitranking:Sampling}. To create this ranking, they collected 121,000 topics from GitHub and selected the top 60\% most frequently occurring topics. They further narrowed this selection down to 3,000 topics based on annotations from three annotators, ultimately extracting 368 software domains. For our study, we chose the top 52 domains, collectively covering 50\% of the repositories, to ensure representation across a substantial portion of OSS domains.



Next, we performed GitHub searches using the GitHub API where for each OSS domain, we used the query "\textit{topic:<OSS Domain> readthedocs.io in:wiki or readthedocs.io in:readme sort:stars}" to locate top-starred repositories that have Sphinx-based documentation hosted on Read the Docs. 
Afterward, we searched Read the Docs for projects with corresponding repository URLs and names to extract their documentation URLs. This process resulted in the collection of 1,387 OSS documentation URLs. To ensure convenient access and minimize reliance on online references, we utilized \textit{HTTrack} to create local mirrors of the documentation for these projects. We were able to mirror 1,383 of the 1,387 documentation websites.


In the following phase, we encountered the challenge of deciding the optimal content granularity to extract from each OSS documentation. Given that BERTopic utilizes Sentence Transformers and acknowledging the significant variability in section and paragraph lengths due to diverse contexts and writing styles, we chose to extract sentences. This decision served two primary purposes: 1) Extracting sentences helped mitigate the problem of padding that could potentially lead to reaching the maximum token limit imposed by the embedding model used in BERTopic. 2) By extracting sentences, we ensured a consistent range of document lengths, thus creating a more manageable and uniform dataset. Our final dataset encompassed 744,608 unique sentences.

\subsubsection{Topic Modeling}

Different variations of topic modeling are supported by BERTopic, such as dynamic, multimodal, hierarchical, online, and semisupervised approaches. Using vanilla BERTopic, even though topics are relevant to one of our PAP-selected criteria, their representations may mislead the topic model towards OSS domain-specific terms. To address this issue and guide the topic model towards creating topics that align more closely with our areas of interest, we can introduce a weighting mechanism for specific words within the dataset. This approach ensures that the resulting topic model contains topics whose representations are more in line with our criteria. BERTopic introduces a semisupervised variation of topic modeling known as Guided Topic Modeling, which facilitates this process by allowing us to guide the model's topic creation using weighted terms called \textit{seed}, thereby producing topics that better match our intended criteria. Seed is a nested list of terms, where each inner list consists of phrases relevant to one of our desired topics. We curated our seed by inspecting OSS documentation pages and extracting common terminology for each criterion, for instance, for the License criterion, we created a list in our seed with the terms: "license", "MIT", "bsd", "GPL", "Apache", "Redistribution", "copyright notice", "disclaimer", "LIABLE", "AS IS", and "MERCHANTABILITY". We tested our seed in three iterations and improved it in each iteration. The seed we used to train the topic model is available at supplementary \cite{Imani2024DoesAdoption}.

\begin{figure}[tp]
  \includegraphics[width=0.8\textwidth]{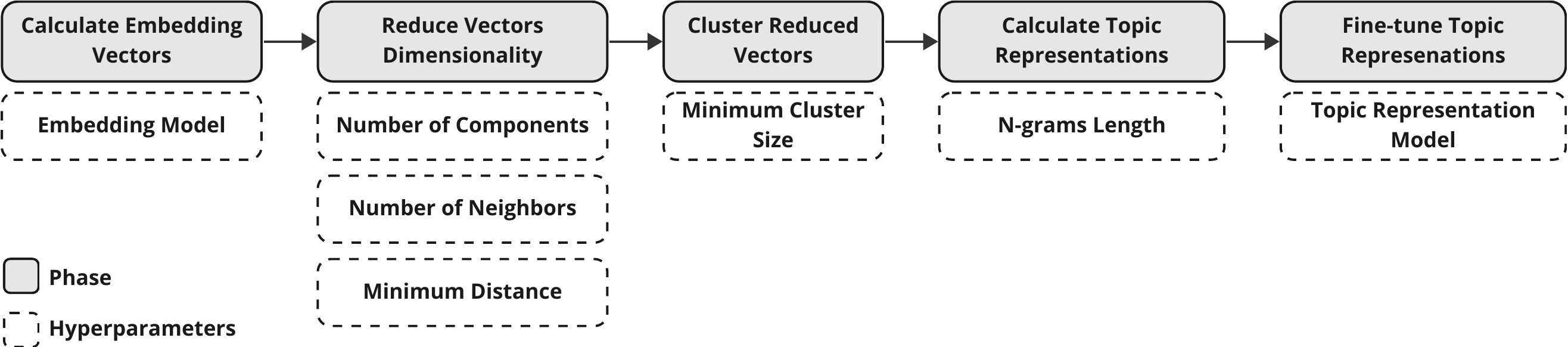}
  \caption{Topic modeling pipeline}
  \label{fig:meth1}
\end{figure}

BERTopic is designed to be modular, meaning each algorithm phase may be customized. As such, ensuring the appropriate configuration at each step is of the utmost importance. Topic modeling phases and their corresponding hyperparameters are summarized in Figure \ref{fig:meth1}.

The first phase in the BERTopic pipeline is calculating embedding vectors for the input documents. Various embedding models are supported in this step. Thus, our first variable that can significantly affect the final topics is the \textbf{embedding model} employed. We focused on the top four SBERT \cite{2019Reimers} pre-trained embedding models known for their robust performance in encoding sentences across a spectrum of 14 distinct tasks. 
However, due to efficiency considerations, we opted to exclude \textit{all-distilroberta-v1} from consideration, as it incurred an 87.5\% increase in processing time compared to \textit{all-MiniLM-L12-v2} with only a negligible 0.03 average performance difference on a magnitude scale of 100.

In the second phase, the dimensionality of the embedding vectors is reduced to counteract the curse of the dimensionality problem in the clustering phase. To handle large amounts of documents, cuML's UMAP implementation can be used to enable GPU utilization in this phase. Although other dimensionality reduction algorithms could be used, we adopted UMAP as its superiority is discussed in previous work \cite{Grootendorst2022BERTopic:Procedure,McInnes2018UMAP:Reduction}. CuML's UMAP offers a range of tunable parameters, but among them, two variables highlighted by BERTopic have a substantial impact on the algorithm: the \textbf{number of neighbors} and the \textbf{number of components}. The number of neighbors, which can be set between 2 and 100, plays a crucial role in determining the preservation of local data. Smaller values favor the retention of more localized data. Additionally, the number of components represents the dimensionality of the reduced vectors. Increasing the value of this variable reduces the impact of dimensionality reduction on clustering. In addition to the aforementioned hyperparameters, there is another crucial parameter to consider in cuML UMAP: the \textbf{minimum distance} between the reduced vectors. Ranging from 0 to 1, this parameter plays a pivotal role in the balance between global and local structure preservation. When set to smaller values, it leads to a more tightly clustered embedding space, where vectors in close proximity on the manifold are drawn even closer together. We initiated with a conservative minimum value of 5 for both hyperparameters. Subsequently, we incremented these values in steps of 5, gradually increasing them to 20. This enabled the exploration of larger values in cases where we observed an upward trend in the topic model's performance. 

In the third phase, reduced vectors are clustered. Similar to previous phases, BERTopic supports customized clustering algorithms. Since not all the information within OSS documentation aligns with practitioners' adoption criteria, it is prudent to employ a clustering algorithm that accommodates outlier data points. One such algorithm, used as the default in BERTopic, is HDBSCAN \cite{McInnes2017Hdbscan:Clustering.}. HDBSCAN simplifies the task by preventing the need to select the number of clusters manually, as well as excelling at automatically clustering dense areas of varying densities. To harness GPU acceleration benefits, we opted for cuML's HDBSCAN implementation.
While this algorithm offers flexibility through various customizable parameters, such as the minimum sample size and the metric used for distance calculations, a crucial parameter to consider is the \textbf{minimum cluster size}. This parameter determines the coherence of the resulting topics. If the minimum cluster size is too high relative to the number of documents, it can lead to the clustering of irrelevant documents, adversely impacting the quality of topic representations. Considering the substantial volume of sentences, numbering in the hundreds of thousands, to ensure coherent topics, we chose a minimum value of 50. It could be argued that this selection might be reduced further, but our experiments indicated that lower values, such as 20, resulted in a significant proliferation of topics, each containing only a minimal number of sentences. Conversely, experimenting with larger values could have led to topic merging, incompatible with our final topic prediction pipeline.

Topic representations are terms extracted from a topic's documents that effectively represent the topic documents. They are pivotal in predicting topics for unseen documents. There are two modules that work together to calculate them: Scikit-learn CountVectorizer and BERTopic's c-TF-IDF. In addition to the CountVectorizer parameters which control the creation of the term-topic matrix used in c-TF-IDF, there are several other customizable aspects to consider. One such aspect is the option to set stop words, enabling the exclusion of common stop words from the topic representations. Furthermore, \textbf{the length of N-grams} to be considered during the c-TF-IDF calculation can be tailored within CountVectorizer. 
We conducted experiments with both unigrams (N-grams of length 1) and bigrams (N-grams of length 2). Through our observations, we discerned that increasing the N-gram length introduced variations of the same word into topic representations. For instance, a topic comprising sentences relevant to compatibility might be represented by both "\textit{install pip}" and "\textit{install conda}". However, reducing the N-gram length to 1 results in "\textit{install}" appearing only once in the topic representation. This, in turn, allows other N-grams to represent the topic, fostering wider diversity in topic representations.

To address the issue of frequent terms that may not be classified as stop words, it is possible to configure the c-TF-IDF algorithm to calculate the square root of term frequency after normalizing the frequency matrix, thereby refining topic representation quality. We enabled the stop word exclusion option in CountVectorizer as well as the frequency reduction option in c-TF-IDF so as to minimize the occurrence of repeating terms in topic representations.

After training the model, the topic representations can be fine-tuned. BERTopic supports various representation models, such as Maximal Marginal Reference, KeyBERT, etc., that can be chained. Experimenting with different \textbf{topic representation models} is crucial since topic models depend heavily on topic representations for valid predictions. 
We conducted experiments using four distinct approaches: Zero-shot classification, \textit{KeyBertInspired} (\textbf{KBI}), \textit{MaximalMarginalRelevance} (\textbf{MMR}) \cite{Carbonell1998TheSummaries}, and a combination of the latter two methods. Notably, zero-shot classification limits the number of terms representing each topic to one. Consequently, it becomes challenging to recognize the underlying themes of sentences within each topic using this approach.

Drawing inspiration from KeyBERT \cite{Grootendorst2020KeyBERT:BERT}, the KBI representation model selects topic representations based on semantic similarity to topic embeddings. These embeddings are computed using representative documents per topic, identified by their c-TF-IDF scores. To enhance the diversity of keywords in topic representations and mitigate redundancy (e.g., the co-occurrence of "\textit{install}" and "\textit{installation}" in a single topic representation), we incorporated the use of MMR. By chaining the MMR and KBI approaches, we aimed to balance preserving semantic relationships and ensuring diversity within the topic representations simultaneously.

For each hyperparameter, H, we froze the values of other hyperparameters and experimented with potential values for H. We randomly sampled 447 sections from our dataset, ensuring a sampling confidence of 95\% with a margin of error of 5\%. Subsequently, the first and second authors independently labeled the samples, and any instances with conflicting labels were designated as outliers. In total, our labeling process yielded an agreement rate of 81\% on the labels. We refer to this validation sample as \textit{Groundtruth}.

\begin{figure}[bp]
  \includegraphics[width=0.8\textwidth]{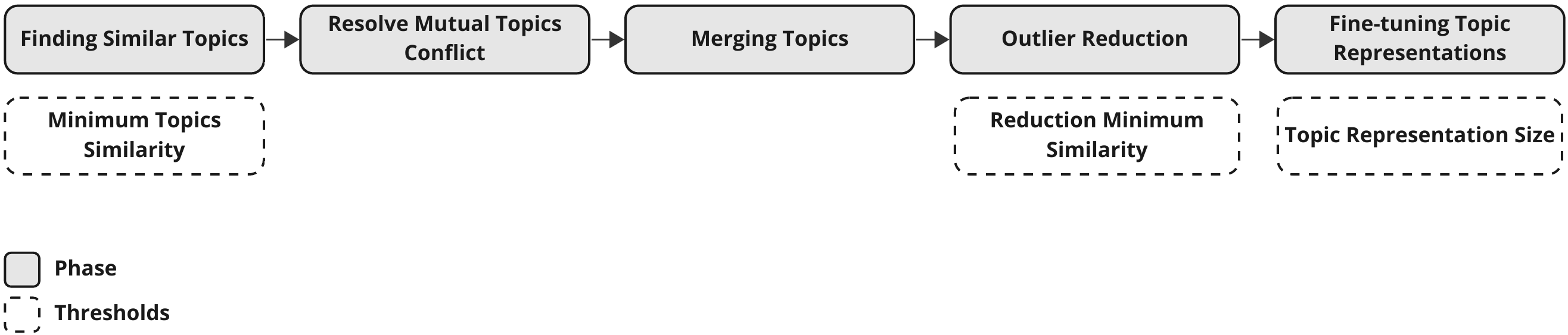}
  \caption{Topic prediction pipeline}
  \label{fig:meth2}
\end{figure}

\subsection{Topics of Interest}
In the process of predicting topics for unseen sentences, BERTopic relies on the calculation of their embedding vectors and measures their cosine similarity to topic embeddings. These topic embeddings are derived by averaging the embeddings of sentences within each topic. To ensure satisfactory performance, it is crucial to maintain minimal similarity between the topics of interest (TOIs). To achieve this, we chose to merge conceptually similar practitioners' criteria, an essential step in mitigating potential mispredictions. For example, topics like "project's versioning" and "project's maintenance" share overlapping terminology, making it challenging to distinguish sentences relevant to these two distinct topics. 
To achieve this objective, thorough discussions were conducted between the first and second authors, resulting in complete consensus. The aim of these discussions was to minimize the merging of criteria while still encompassing practitioners' preferences effectively. Ultimately, the merged criteria were closely aligned with recent developments in OSS quality factors \cite{Madaehoh2022OSS-AQM:Measurement}.


\begin{itemize}
    \item \textbf{License}: License Compatibility
    \item \textbf{Functional Suitability}: Ease of use, Functionality \& Features, The existence of examples of use cases, Project's Performance
    \item \textbf{Compatibility}: Compatibility
    \item \textbf{Project's Maintenance}: Project's Maintenance, Project's usage trends, Project's versioning, Community Adoption
\end{itemize}

Finally, we considered a topic called \textbf{Outlier} to include sentences irrelevant to the aforementioned TOIs.

\subsection{Topic Prediction}

Due to the substantial size of our sentence dataset, which comprises 744,608 sentences from diverse OSS domains, BERTopic naturally generates a large number of topics. However, our focus is on identifying five specific topics: License, Functional Suitability, Compatibility, Project Maintenance, and an outlier topic. 
To achieve this, we employ a strategy to identify topics similar to each of our TOIs and subsequently merge them. It might be argued that combining these topics is unnecessary, as they could simply be marked as one of the TOIs. However, based on our experiments, not merging the topics leads to an increased number of false positive predictions for the outlier topic. By merging similar topics, we create larger composite topics, and their embeddings are calculated through a weighted average of the embeddings of the initially merged topics. This strategy enhances the likelihood of correctly assigning an unseen sentence to the appropriate topic. It considers both the original topics' sizes and their embeddings, contributing to improved topic prediction performance.

\begin{figure}[hbp]
\centering
\pgfplotstableread{
0	73.79	73.58	60
1	50.5	51.46	51.89
2	84.96	87.27	91.59
3	72.34	72.73	70.21
4	69.37	69.79	70.71
5	66.52	68.2	67.38

}\database
\begin{tikzpicture}[
every axis/.style={
      width  = 0.7*\textwidth,
      height = 6cm,
      major x tick style = transparent,
      ymin=0,
      scaled y ticks = false,
      ymin=0,
      ymax=140,
      bar width=.5em,
        ybar,
        ylabel={Median Weighted F1-Score},
        ylabel style={ font=\scriptsize\bfseries},
        xtick=data,
        ytick = {0,20,40,60,80,100},
        xticklabels = {
            Project's Maintenance,
            Functional Suitability,
            License,
            Compatibility,
            Outlier,
            Overall,
        },
        xticklabel style={rotate=30, anchor=center, font=\scriptsize\bfseries,yshift=-1em,text width=2cm, align=center,xshift=-2em},
        major x tick style = {opacity=0},
        minor tick length=2ex,
          legend cell align=center,
          legend columns=-1,
      legend style={
          column sep=1ex,
          font=\tiny\bfseries
        },
}
]
\pgfplotsset{compat=1.11,
    /pgfplots/ybar legend/.style={
    /pgfplots/legend image code/.code={%
       \draw[##1,/tikz/.cd,yshift=-0.25em]
        (0cm,0cm) rectangle (5pt,0.8em);},
   },
}
\begin{axis}
\addplot[fill=black,postaction={
        pattern=horizontal lines
    }] table[x index=0,y index=1] \database; 
\addplot[postaction={
        pattern=horizontal lines
    }] table[x index=0,y index=2] \database; 
\addplot[postaction={
        pattern=crosshatch
    }] table[x index=0,y index=3] \database; 

\legend{
all-MiniLM-L12-v2,
all-MiniLM-L6-v2,
all-mpnet-base-v2
}
\end{axis}

 \end{tikzpicture}
   \caption{Embedding models prediction performance comparison. The Y values present the median of the embedding model's weighted average F1-score within all hyperparameters.}
  \label{fig:embedding}
\end{figure}
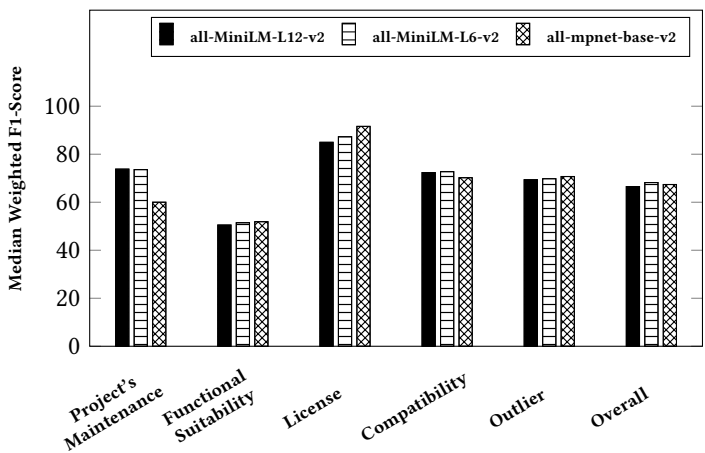



Our topic prediction pipeline is depicted in Figure 3. It commences with the identification of topics similar to our seed topics by concatenating the phrases (search strings) within each TOI and computing the cosine similarity between the search string and topic embeddings. If the similarity surpassed or equaled the \textit{Topics Similarity Threshold}, we added the topic and its similarity score to the TOI's merge list. Since certain topics appeared in the merge lists of multiple TOIs, we resolved conflicts over mutual topics by removing the topic from the TOI with the lower cosine similarity to it. Subsequently, we merged the topics and labeled them with their corresponding TOI names. Among the resulting merged topics, the outlier topic was the largest, which increased the risk of false positive predictions for this topic. To address this, we reduced the outliers by assigning their sentences to one of the TOIs based on the \textit{Reduction Minimum Similarity} criterion. The topic representations of the merged topics were updated through a combination of MMR and KBI. For each final topic, the \textit{Topic Representation Size} parameter determined the number of terms included in the topic representations.


Our final topic predictor operates by receiving a section, predicting the topics for the sentences within it, and subsequently using the total similarity values between each sentence and the topics to make a prediction regarding the section's topic. While alternative methods, such as calculating the median or mean of sentence similarities to the topics, could be employed, our experimental findings have shown that summing the similarities consistently leads to more accurate predictions at the section level.

Accordingly, our topic prediction pipeline introduces three thresholds: Topics Similarity Threshold, Reduction Minimum Similarity, and Topic Representation Size. For the first two thresholds, we conducted an extensive search, ranging from 0 to 1 with increments of 0.1. Additionally, we explored values within the range of 10 to 50 for Topic Representation Size, with increments of 10. 

We discuss optimal hyperparameters performance regarding the weighted average F1 score on all our TOIs. We chose the F1 score to take into account precision and recall simultaneously. The reason behind selecting a weighted average is to address the class imbalance in Groundtruth. This is because both of these metrics are critical to our topic predictor. 
Regarding the embedding model, all-MiniLM-L6-v2, the default model suggested by BERTopic, outperformed other embedding models in predicting three out of four TOIs. However, all-mpnet-base-v2 outperformed the others in predicting license-related and outlier sections. Considering overall performance, all-MiniLM-L6-v2 outperformed the other embedding models with a slight superiority of 1\%. The median weighted average F1-score across all the TOIs achieved by all-MiniLM-L6-v2 is 67.4\%. Figure \ref{fig:embedding} illustrates embedding models' performance across TOIs.

\begin{figure}[tp]
  \includegraphics[width=0.7\textwidth]{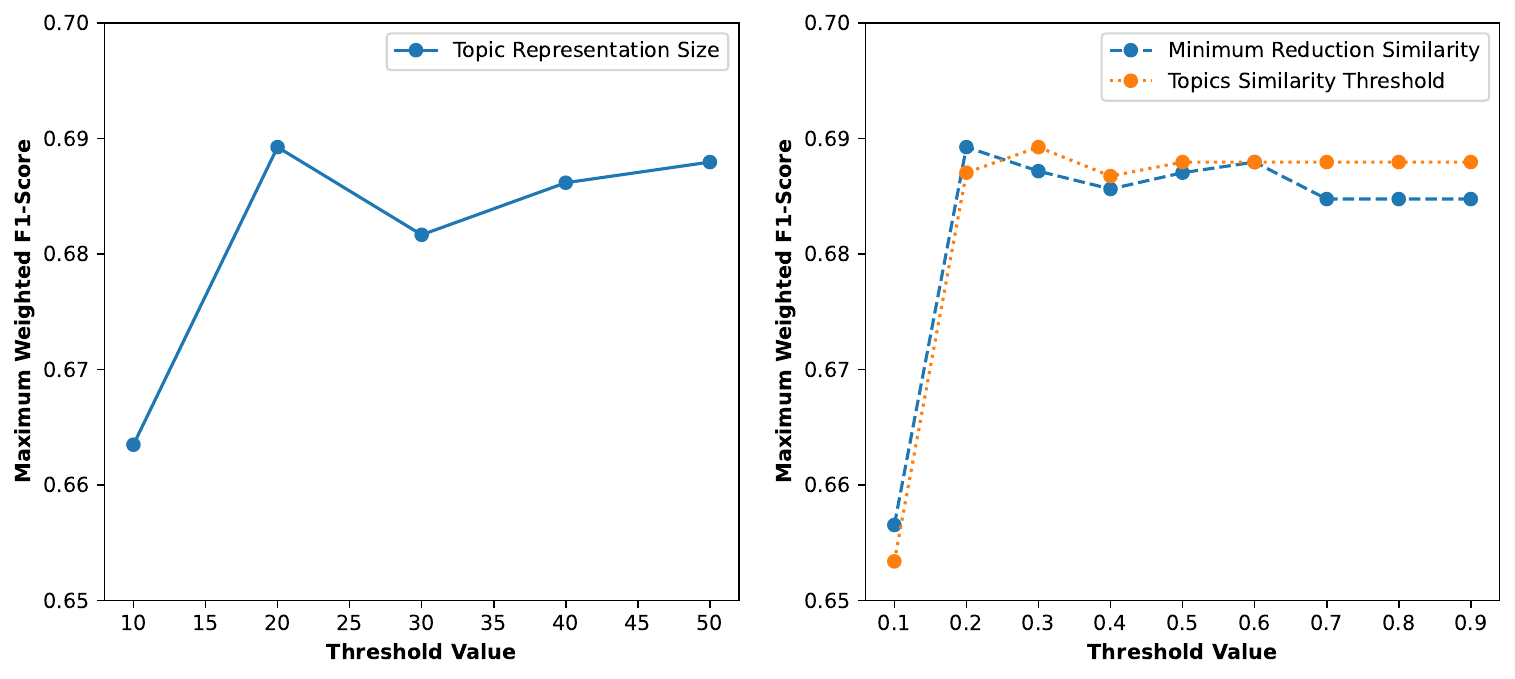}
\setlength{\belowcaptionskip}{-6mm}
  \caption{Topic prediction thresholds performance comparison. Note that the Y ranges have been limited to a range that eases observing the changes.}
  \label{fig:thresholds}
\end{figure}

Setting the embedding model to all-MiniLM-L6-v2, we experimented with the rest of the hyperparameters, the number of components, neighbors, and the minimum distance between reduced embedding vectors. Changes in none of the hyperparameters impacted the median weighted average F1-score of all-MiniLM-L6-v2 to the magnitude of 0.001s. However the optimal values for these hyperparameters that maximized the weighted average F1-score were 20, 20, and 0.1. Figure \ref{fig:thresholds} presents the maximum weighted average F1 score achieved by each threshold value. Topic Representation Size's performance peaked at 20, the Minimum Reduction Similarity at 0.2, and the Topics Similarity Threshold at 0.3. 


\subsection{DocMentor}

In order to enable practitioners with different OSS domain knowledge levels, we propose DocMentor which combines TF-IDF and ChatGPT to identify technical terms and provide explanations, references, and examples. Following, we analyze TF-IDF's application for technical terms detection. \

Considering the broad spectrum of OSS domains, technical terms specific to an OSS domain, $D_i$, may have a low frequency of occurrence in documents of other OSS domains. When calculating TF-IDF, these terms will have a low term frequency (\textbf{TF}) in non-relevant OSS domains, contributing to a higher TF-IDF score when the term is present in $D_i$. Conversely, OSS domain-specific terms often exhibit high document frequency (\textbf{DF}) within their respective domains. For example, terms like "f1-score" may appear frequently in numerous OSS documentation sections within the Machine Learning domain. Consequently, this elevated DF translates to a high Inverse Document Frequency (\textbf{IDF}), thereby inflating the TF-IDF scores for technical terms within their specific OSS domains. Accordingly, in the context of TF-IDF, a technical term is identified by three characteristics: 1) High intradomain DF  2) High intradomain TF 3) Low interdomain DF.


While our TF-IDF-based approach effectively captures known technical terms within our corpus, it encounters limitations when identifying unseen technical terms. To optimize the efficacy of our method, complementary approaches become essential. In this regard, we leveraged OpenAI ChatGPT, a state-of-the-art large language model. ChatGPT was harnessed not only to enhance our technical term identification performance but also to provide explanations for the identified technical terms, references, and examples. It is worth noting that ChatGPT has demonstrated its utility in similar tasks, including code explanation \cite{Chen2023GPTutor:Explanation}, making it a valuable asset in DocMentor.

Accordingly, DocMentor's workflow commences with inputting a text segment, such as a paragraph. In this initial stage, technical terms are detected utilizing TF-IDF. Subsequently, leveraging the capabilities of OpenAI's ChatGPT, the list of technical terms is further expanded. Finally, ChatGPT provides brief explanations, and depending on the context, examples, and references for the identified technical terms.
To arrive at a suitable prompt for each interaction with ChatGPT, the first author engaged in an iterative process, guided by our predefined evaluation criteria. Once the final prompts were established, all authors collectively reviewed ChatGPT's responses in a pilot study before subjecting them to evaluation by practitioners. 
The supplementary \cite{Imani2024DoesAdoption} provides an example of DocMentor's input, our ChatGPT prompts and the generated output.

In order to evaluate DocMentor's performance, we initially explored the availability of existing benchmarks that could facilitate a systematic evaluation. However, to the best of our knowledge at the time of this study, no such benchmark existed for assessing augmented information within the domain of OSS.
Hence, we devised an evaluation survey. To ensure an unbiased representation of various criteria, we incorporated one paragraph from each practitioner's criteria. This approach was chosen to maintain manageability for practitioners, as evaluating multiple paragraphs for each criterion would have been impractical. To achieve this, we employed our trained topic model to predict the topics of sections within our OSS documentation corpus. Subsequently, we randomly sampled medium-length paragraphs from each category. If, by chance, the initially selected paragraph was misclassified, we replaced it with another randomly selected paragraph. The resulting set of selected paragraphs was then processed through DocMentor, with the outputs from DocMentor being collected for evaluation through the survey. 
In order to ensure that our survey questions effectively align with the goal of evaluating DocMentor's responses from multiple perspectives, researchers conducted a pilot study and iteratively refined the questions. This iterative process aimed to enhance the clarity and precision of the survey questions,
Our evaluation survey structure is outlined below.

In the first three questions, we sought participants' demographic information. Following that, for each of our TOI-relevant paragraphs, we asked participants to assign a 5-point scale Linkert score to the augmented information on various aspects: 1) its informativeness, 2) The degree to which it clarified technical terms within the paragraph, 3) The relevance of the provided examples or references, 4) The extent to which it would aid in facilitating adoption decision-making, and 5) The extent to which it improved the participant's comprehension of the paragraph.


In order to distribute the survey, we contacted the initial survey and interview participants who opted for participating a potential follow-up survey. To ensure an adequate number of participants, two individuals were invited separately by the researchers, as not all potential survey participants were able to take part in the evaluation survey. 













\section{Results}

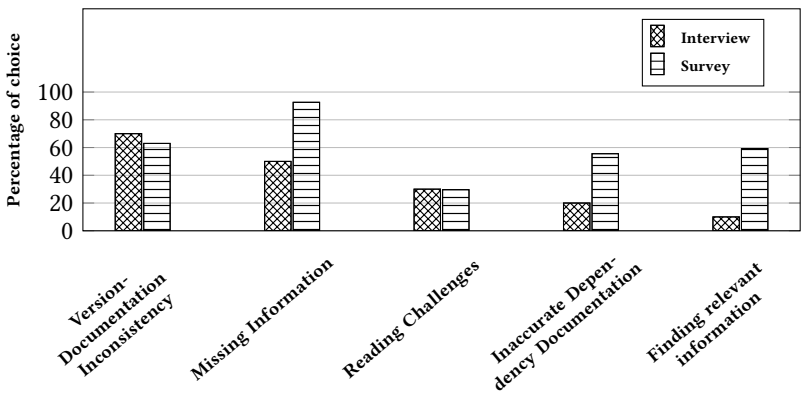
\begin{figure}[bp]
\centering
\begin{tikzpicture}
\pgfplotsset{compat=1.11,
    /pgfplots/ybar legend/.style={
    /pgfplots/legend image code/.code={%
       \draw[##1,/tikz/.cd,yshift=-0.25em]
        (0cm,0cm) rectangle (5pt,0.8em);},
   },
}
  \begin{axis}[
      width  = 0.8*\textwidth,
      height = 4.5cm,
      major x tick style = transparent,
      ybar=2*\pgflinewidth,
      bar width=10pt,
      ymajorgrids = true,
      ylabel = {Percentage of choice},
      ylabel style={ font=\scriptsize\bfseries, yshift = -.2em},
      symbolic x coords={Version-Documentation Inconsistency,Missing Information,Reading Challenges, Inaccurate Dependency Documentation, Finding relevant information},
      xtick = data,
      scaled y ticks = false,
      ymin=0,
      ymax=160,
      ytick = {0, 20, 40, 60, 80, 100},
      x tick label style={
          rotate=40,
          anchor=east,
          text width=2.5cm,
          align=center,
          font=\scriptsize\bfseries,
          xshift = 12pt,
          yshift = -15pt
        },
      legend cell align=left,
      legend style={
          at={(0.95,0.65)},
          anchor=south east,
          column sep=1ex,
          font = \tiny
        }
    ]
    \addplot[style={postaction={pattern=crosshatch}}]
    coordinates {(Version-Documentation Inconsistency, 70) (Missing Information,50) (Reading Challenges,30) (Inaccurate Dependency Documentation, 20) (Finding relevant information, 10)};

    \addplot[style={postaction={pattern=horizontal lines}}]
    coordinates {(Version-Documentation Inconsistency, 62.96) (Missing Information,92.59) (Reading Challenges,29.62) (Inaccurate Dependency Documentation, 55.55) (Finding relevant information, 59.25)};

    \legend{\textbf{Interview}, \textbf{Survey}}
  \end{axis}
\end{tikzpicture}
\caption{Challenges with using OSS documentation}
\label{fig:challenges}
\end{figure}

\subsection{Interview and Survey Insights}
Our findings indicate that within the industry, OSS documentation is regarded as a significant resource for making decisions concerning the adoption of open-source software.
All ten participants confirmed that they rely on OSS documentation when making decisions regarding the adoption of open-source projects into their products. A significant majority of our survey participants, specifically 91.18\%, indicated that they rely on open-source software documentation when making decisions about integrating them into their products.

\begin{table}[tp]
\footnotesize
\setlength{\tabcolsep}{3pt}
  \caption{Participants' Confidence Level in Finding Relevant Criteria in OSS Documentation}
  \label{tab: confidence-level}
  \begin{tabularx}{\linewidth}{lXX}
    \toprule
    Adoption Category& Median of Confidence Level (Interview)& Median of Confidence Level (Survey)\\
    \midrule
    Project Maintenance&4&3\\
    Compatibility&4&3\\
    Functionality&3&4\\
    License Compatibility&3&5\\
    Community Adoption&4&3\\
    Examples of Use Cases&3&4\\
    Ease of use&4&3\\
    Project's Versioning&2.75&3.5\\
    Performance&4&2.5\\
    \bottomrule
   \end{tabularx}
\end{table}

When asked to rate the importance of OSS documentation in their decision on a scale of 1 to 5, with 5 being the highest, the median of interview responses was 5. Specifically, one participant rated it as 2, another as 3, two participants as 4, and six participants as the highest rating, 5. The median of the survey participants'  responses to this question was 4. More specifically, 3.7\% rated it as 1, 22.22\% as 2, 37.04\% as 4, and another 37.04\% as 5. This indicates a strong consensus among the participants regarding the significant role of OSS documentation in their decision-making process. Table \ref{tab: reason} displays the survey participants' responses regarding the reasons for utilizing open-source software documentation when making decisions about adopting OSS.



In Figure \ref{fig:challenges}, we present the challenges reported by the participants when using OSS documentation. Each challenge is accompanied by the corresponding percentage, indicating the proportion of participants who mentioned that issue when using OSS documentation.

\begin{figure}[tp]
    \makeatletter
\pgfplotsset{
    calculate offset/.code={
        \pgfkeys{/pgf/fpu=true,/pgf/fpu/output format=fixed}
        \pgfmathsetmacro\testmacro{(\pgfplotspointmeta *10^\pgfplots@data@scale@trafo@EXPONENT@y)*\pgfplots@y@veclength)}
        \pgfkeys{/pgf/fpu=false}
    },
    every node near coord/.style={
        /pgfplots/calculate offset,
        yshift=-\testmacro-15pt
    },
}
\pgfplotstableread{
0 30 20 30 0 20
1 30 30 0 0 40
2 30 10 20 0 40
3 40 0 0 0 60
4 10 10 30 0 50
5 10 10 10 0 70
6 10 0 10 10 70
7 10 0 10 0 80
8 0 10 0 0 90
}\interview
\pgfplotstableread{
0 20.68 0 10.34 27.58 41.4
1 3.44 13.79 10.34 10.34 62.09
2 20.68 34.48 10.34 6.89 27.61
3 31.03 3.44 3.44 10.34 51.75
4 0 10.34 20.68 27.58 41.4
5 3.44 6.89 0 27.58 62.09
6 3.44 13.79 17.24 17.24 48.29
7 0 0 0 3.44 96.56
8 0 0 6.89 24.13 68.98
}\survey
\begin{tikzpicture}[
every axis/.style={
      width  = 0.8*\textwidth,
      height = 7cm,
      major x tick style = transparent,
      ymin=0,
      scaled y ticks = false,
      ymin=0,
      ymax=140,
      bar width=.5em,
        ybar stacked,
        ylabel={Percentage of choice},
        ylabel style={yshift=-0.4cm, font=\scriptsize\bfseries},        
        xtick=data,
        ytick = {0,20,40,60,80,100},
        xticklabels = {
            Maintenance,
            Compatibility,
            Functionality,
            License,
            Community Adoption,
            Examples of Use Cases,
            Ease of use,
            Project’s Versioning,
            Performance,
        },
        xticklabel style={yshift=-4ex, rotate=30, anchor=east, font=\tiny\bfseries},
        major x tick style = {opacity=0},
        minor tick length=2ex,
          legend cell align=center,
          legend columns=2,
      legend style={
          column sep=1ex,
          font=\tiny
        },
}
]
\begin{axis}[reverse legend]
\addplot[fill=black,postaction={
        pattern=north east lines
    }] table[x index=0,y index=5] \interview; 
\addplot[postaction={
        pattern=north west lines
    }] table[x index=0,y index=4] \interview; 
\addplot[postaction={
        pattern=crosshatch
    }] table[x index=0,y index=3] \interview; 
\addplot[postaction={
        pattern=dots
    }] table[x index=0,y index=2] \interview; 
\addplot[nodes near coords=I,postaction={
        pattern=horizontal lines
    }] table[x index=0,y index=1] \interview;

\legend{
\textbf{Not a Priority},
\textbf{$+4^{\text{th}}$ Priority},
\textbf{$3^{\text{rd}}$ Priority},
\textbf{$2^{\text{nd}}$ Priority},
\textbf{$1^{\text{st}}$ Priority}
}
\node[draw,yshift=-5pt, xshift=-15pt, anchor=north west, font=\footnotesize,align=left] at (axis cs: 0,140) {\textbf{S: Survey}\\ \textbf{I: Interview}};

\end{axis}

\begin{axis}[bar shift=-8pt,hide axis]
\addplot[fill = black, postaction={
        pattern=crosshatch
    }] table[x index=0,y index=5] \survey; 
\addplot[postaction={
        pattern=north west lines
    }] table[x index=0,y index=4] \survey; 
\addplot[postaction={
        pattern=crosshatch
    }] table[x index=0,y index=3] \survey; 
\addplot[postaction={
        pattern=dots
    }] table[x index=0,y index=2] \survey; 
\addplot[nodes near coords=S, nodes near coords style={xshift=-8pt}, postaction={
        pattern=horizontal lines
    }] table[x index=0,y index=1] \survey;

\end{axis}
 \end{tikzpicture}
 \makeatother
    \caption{Importance of Adoption Metrics in OSS Documentation}
    \label{fig:importance}
\end{figure}
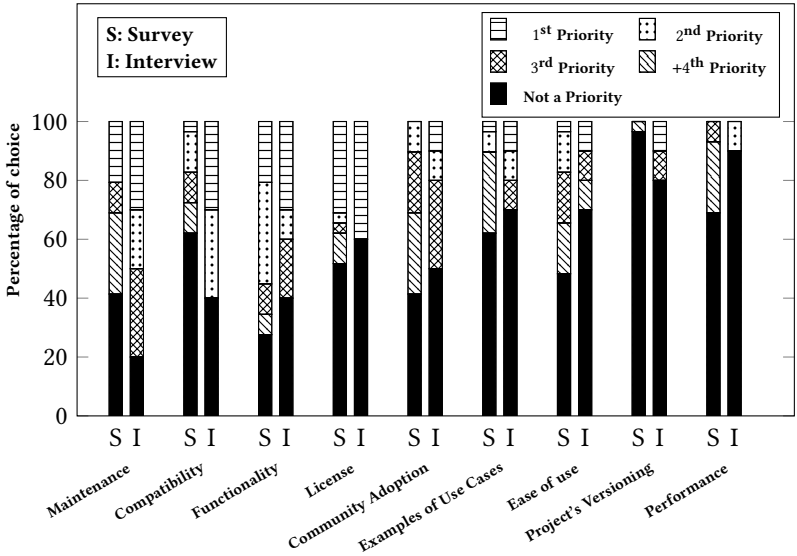

To assess the importance of each criterion, participants were asked to rank them based on their importance. Figure \ref{fig:importance} illustrates the relative importance of each category as perceived by the participants. Table \ref{tab: confidence-level} displays the participants' confidence levels for each adoption criterion when seeking relevant information in OSS documentation. They were asked to rate their confidence level, ranging from 1 to 5, with 5 indicating the highest level of confidence. 



\begin{table}[tp]
\footnotesize
\setlength{\tabcolsep}{3pt}
  \caption{Participants' Reasons for using open-source documentation during decision-making for OSS adoption }
  \label{tab: reason}
  \begin{tabular}{lc}
    \toprule
    Reason for Using OSS Documentation& Choice Percentage\\
    \midrule
    Implement a proof of concept&10.60\%\\
    Understand how the open-source software works&11.92\%\\
    Realize the open-source software functionalities and features&16.56\%\\
    Analyze how the open-source software can be used&11.26\%\\
    Get introduced to the open-source software&7.28\%\\
    Project's Versioning&2.75\%\\
    Analyze the open-source software community and contributors&5.30\%\\
    Monitor the open-source software maintenance&3.31\%\\
    Study examples of using the open-source software&12.58\%\\
    Verify compatibility (Including license compatibility)&8.61\%\\
    Analyze the open-source software maturity&8.61\%\\
    Understand the contribution procedure&3.31\%\\
    \bottomrule
   \end{tabular}
\end{table}

\subsection{PAP-relevant Information Gathering Performance}

Answering \textbf{RQ3}, using threshold values and the optimal hyperparameters (Explained in Section \hyperlink{section.3}{3}), resulted in a weighted average of 73\% precision, 68\% recall, and 69\% F1-score.

The performance of BERTopic hinges on the cosine similarity between the embeddings of unseen sentences and the topic embeddings. However, this approach introduces the potential for false positives, where sentences may be misclassified into topics they do not belong to. For example, sentences explaining the functional suitability of an OSS library in the context of web servers might include terms like "permission" and "license," which also have high c-TF-IDF values for License-related sentences. This overlap in terminology can result in false predictions for both Topics of Interest (TOIs), namely License and Functional Suitability, underscoring the challenge of disambiguating topics with shared terminology. This prompts the need for future research to alleviate this challenge.

\subsection{DocMentor Evaluation Results}

Our evaluation survey was completed by 10 practitioners, whose demographic information is available at our supplementary \cite{Imani2024DoesAdoption}. Following, we discuss responses to each evaluation aspect as we discussed in Section \hyperlink{section.3}{3}.

This question focused on evaluating ChatGPT's explanations of the technical terms identified by DocMentor. It received a median score of 3.75 when assessed across all paragraphs. Notably, the compatibility paragraph achieved the highest median score of 4, signifying that practitioners found the output particularly informative in this context. Conversely, the output for the license paragraph was rated as less informative, with a median score of 3.50. Several factors may have contributed to this observation. It is possible that the practitioners possessed a high level of domain knowledge regarding licensing terms, which could have influenced their perception of DocMentor's output's informativeness. Alternatively, there might have been issues with accurately identifying technical terms within the paragraph by DocMentor, or the explanations provided may not have sufficient information. Further investigation may be warranted to pinpoint the specific reasons behind these informativeness scores differences.
    
In terms of the outputs' ability to clarify technical terms, the survey revealed a median score of 3.75 when considering all paragraphs collectively. Interestingly, when evaluating individual paragraphs, the  project maintenance-related-related paragraph received the highest score in this regard, with a median of 4. In contrast to the previous question, the compatibility-related paragraph obtained the lowest median score of 3.5 in clarifying technical terms. This intriguing divergence implies that while the output for the compatibility paragraph may have been informative, it fell short in providing the necessary clarification for technical terms. This discrepancy could be attributed to the nature of the technical terms themselves, which may not have aligned with practitioners' perceptions or expectations. 
    
Among the outputs evaluated, the functional suitability paragraph was the sole instance where examples were provided. Participants assessed the relevance of these examples, resulting in a median score of 4. This observation underscores the potential value of leveraging ChatGPT to generate quality OSS-specific examples.
    
Based on the feedback provided by participants on the potential DocMentor's helpfuless in PAP, which yielded a median score of 3, it is evident that there is room for further tailoring DocMentor to better align with the specific needs of the adoption procedure. Notably, practitioners scored the minimum score for the project maintenance paragraph. The participants' evaluations have raised important questions and considerations, prompting the need for a more in-depth investigation. This further exploration aims to gain a deeper understanding of how DocMentor's responses can be optimized to effectively facilitate PAP and cater to the unique requirements and expectations of practitioners.
    
Practitioners' understanding of the paragraphs was improved by achieving a median score of 3.75 across all paragraphs. Consistent with previous question findings, practitioners' understanding of the project's maintenance paragraph was minimally impacted by DocMentor's output. This outcome was anticipated, as project maintenance terminology (such as version control-relevant terms) tends to be familiar to practitioners with even minimal domain knowledge. In contrast, DocMentor's impact on practitioners' comprehension of the license-related paragraph was more pronounced, reaching a median score of 4. This suggests that DocMentor played a notably helpful role in empowering practitioners to better understand license-related information during the PAP.

The evaluation survey has unveiled the intrinsic capabilities of DocMentor's augmented information, providing valuable insights into its potential benefits. However, as the inaugural endeavor in augmenting information to facilitate PAP, it is evident that DocMentor should undergo further enhancements to make it more applicable and effective in real-world scenarios. Addressing \textbf{RQ4}, the combination of TF-IDF scores calculated from an OSS documentation corpus with the capabilities of large language models has the potential to enhance practitioners' comprehension of OSS documentation. This approach enables practitioners with diverse levels of domain knowledge to better grasp the information contained within this resource.


\section{Threats to Validity}

We have adopted measures to mitigate potential threats to the validity of our work, which we outline in this section.

\subsection{Construct validity} To mitigate the potential threat to construct validity stemming from participant misunderstandings of our survey questions, we took proactive measures by conducting pilot studies. 
Subsequently, we used the feedback from these pilot studies to refine and update our surveys. This iterative process allowed us to enhance the clarity and comprehensibility of our survey questions.

\subsection{Internal validity} In this study, we utilized a snowball sampling approach for participant recruitment in the interview phase. Although snowball sampling is inherently non-random and can introduce selection bias, we implemented measures to address this potential bias. Initially, we intentionally identified a diverse group of key informants with relevant insights into the research topic. To mitigate bias related to the initial informants, we encouraged them to suggest additional participants from their networks who had experience with software development and OSS adoption experiments. Additionally, we ensured sample diversity by including participants from various industrial sectors to capture a range of perspectives on OSS adoption.

\subsection{External validity} 
To create our OSS documentation dataset, we specifically concentrated on Sphinx-based documentation hosted on Readthedocs. We made this decision because such documentation follows a standardized structure and consistently uses common tags, making data extraction more manageable. Nevertheless, we were aware of the potential limitation of this approach in terms of source diversity. To mitigate this concern, we strategically diversified the OSS domains in our dataset. By including a broad range of OSS domains, we effectively increased the inclusivity and comprehensiveness of our dataset.

\section{Conclusion and Future Work} 
In this study, we investigated the essential role of OSS documentation in practitioners' adoption processes through interviews and surveys. Our findings not only highlight the significance of OSS documentation in adoption but also offer insights into the criteria practitioners use for their adoption decisions.
From the data we collected, we introduced a novel automated method utilizing BERTopic to extract PAP-relevant information. This method lays the foundation for creating a web service that can take an OSS library name as a query and provide well-categorized OSS documentation, customized for the adoption process. As ours was the initial step toward automating the retrieval of OSS adoption-related information, future work can aim to enhance the performance further. Another natural progression of this work would be introducing a personalized OSS adoption assistant based on Natural Language Inference. This assistant could analyze adoption criteria and produce compliance reports, reducing human judgment and the likelihood of errors in the adoption process.

\section{Data Availability}

Our study's data, including interview scripts, the initial and evaluation survey results, all the Python code for different sections, our trained topic model, and the DocMentor's code along with its ChatGPT prompts, are available at supplementary \cite{Imani2024DoesAdoption}.

\bibliographystyle{ACM-Reference-Format}
\bibliography{main}










\end{document}